\documentclass[fleqn,10pt]{wlscirep}
\pdfoutput=1
\usepackage[utf8]{inputenc}
\usepackage[T1]{fontenc}
\usepackage{outlines}
\usepackage{booktabs}
\usepackage{color, colortbl}
\usepackage{tabularx}
\usepackage{mathtools}
\usepackage{amssymb}
\usepackage{amsmath}
\usepackage{textcomp}
\usepackage{multirow}
\usepackage{titlesec} 
\usepackage{floatrow}
\usepackage{setspace}
\usepackage{algorithmicx}
\usepackage{multicol}
\usepackage{csvsimple}
\usepackage{algpseudocode}
\usepackage[superscript,biblabel]{cite}
\usepackage{pgfplotstable}
\usepackage{array}
\usepackage{filecontents} 
\usepackage{subcaption}
\usepackage{todonotes}
\usepackage{authblk}
\usepackage{colortbl}
\usepackage{graphicx}
\pgfplotsset{compat=1.16,
/pgfplots/ybar legend/.style={
/pgfplots/legend image code/.code={
\draw [##1,/tikz/.cd,bar width=4pt,yshift=-0.2em,bar shift=0pt]
plot coordinates {(0cm,0.8em)};
}
}}
\usetikzlibrary{patterns}
\usepackage{tikz}
\pgfplotsset{
    discard if/.style 2 args={
        x filter/.code={
            \edef\tempa{\thisrow{#1}}
            \edef\tempb{#2}
            \ifx\tempa\tempb
                
            \fi
        }
    },
}
\hypersetup{hidelinks}
\definecolor{Gray}{gray}{0.9}
\newcommand{\ra}[1]{\renewcommand{\arraystretch}{#1}}

\newcommand{\norm}[1]{\left\lVert#1\right\rVert}

\newlength{\adaptivefigwidth}
\begin{abstract}
Advances in imaging and early cancer detection have increased interest in magnetic resonance (MR) guided focused ultrasound (MRgFUS) technologies for cancer treatment. 
MRgFUS ablation treatments could reduce surgical risks, preserve organ tissue/function, and improve patient quality of life.
However, surgical resection and histological analysis remain the gold standard to assess cancer treatment response.
For non-invasive ablation therapies such as MRgFUS, the treatment response must be determined through MR imaging biomarkers. 
However, current MR biomarkers are inconclusive and have not been rigorously evaluated against histology via accurate registration.
Existing registration methods rely on anatomical features to directly register \textit{in vivo} MR and histology.
For MRgFUS applications in anatomies such as liver, kidney, or breast, anatomical features independent from treatment features are often insufficient to perform direct registration. 
We present a novel MR to histology registration workflow that utilizes intermediate imaging and does not rely on these independent features.
The presented workflow yields an overall registration accuracy of 1.00 +/- 0.13 mm.
The developed registration pipeline is used to evaluate a common MRgFUS treatment assessment biomarker against histology.
Evaluating MR biomarkers against histology using this registration pipeline will facilitate validating novel MRgFUS biomarkers to improve treatment assessment without surgical intervention.

\end{abstract}

\title{\Large Histology to 3D \textit{In Vivo} MR Registration for Volumetric Evaluation of MRgFUS Treatment Assessment Biomarkers}%
\author[1,2*]{Blake~E.~Zimmerman}
\author[1,3]{Sara~L.~Johnson}
\author[3]{Henrik~A.~Od\'een}
\author[4]{Jill~E.~Shea}
\author[5]{Rachel~E.~ Factor}
\author[1,2]{Sarang~C.~Joshi}
\author[3]{Allison~H.~Payne}
\affil[1]{Department of Biomedical Engineering, University of Utah, Salt Lake City, UT}
\affil[2]{Scientific Computing and Imaging Institute, University of Utah, Salt Lake City, UT}
\affil[3]{Utah Center for Advanced Imaging Research, University of Utah, Salt Lake City, UT}
\affil[4]{Department of Surgery, University of Utah, Salt Lake City, UT}
\affil[5]{Department of Pathology, University of Utah, Salt Lake City, UT}
\affil[*]{Corresponding Author: Blake E. Zimmerman, blakez@sci.utah.edu}
\begin{document}
\flushbottom
\maketitle%
\vspace{-7mm}
\vspace{-1em}
\section{Introduction}

Improved early cancer detection has driven the development of more conservative, less invasive cancer treatment alternatives to surgical intervention.
These minimally and non-invasive tumor treatments have the potential to reduce or eliminate surgical risks, preserve organ tissue and function, and improve patient quality of life.
Magnetic resonance (MR) guided focused ultrasound (MRgFUS) ablation is a rapidly growing technology for non-invasive tumor treatment with applications including liver \cite{kennedy2004high}, prostate \cite{colombel2004principles}, and breast tumors \cite{merckel2018eligibility,sun2014therapies,kaiser2008mri}. 
In any ablative treatment, gold-standard determination of treatment response requires tissue resection and histological analysis \cite{pichat2018survey}. 
However, one of the key benefits of  minimally and non-invasive ablation treatments is the elimination of traditional surgical intervention. 
Therefore, in non-invasive ablation therapies such as MRgFUS, treatment response must be determined through MR imaging biomarkers alone.
Before MR biomarkers can be used to assess MRgFUS treatment response, they must be validated against histology to determine accuracy \cite{futureneedsHectors2016}. 
The most robust way to validate MR biomarkers is through spatially accurate and precise comparison between \textit{in vivo} MR biomarkers and histology. 

Directly comparing MR images of \textit{in vivo} tissue (\textit{in vivo} MR) to histology is challenging. 
Excising treated tissue, slicing into smaller tissue blocks, and preparing tissue samples for histology all destroy the spatial relationship between \textit{in vivo} MR and histology images \cite{gibson20133d}. 
Because of this difficulty, previous studies have avoided restoring this spatial relationship and evaluate MRgFUS MR biomarkers via qualitative or indirect metrics, such as non-viable tumor fractions that can be derived relative to the MR and histology spaces independently \cite{hectors2014multiparametric}. 
Qualitative analysis and indirect metric evaluation have been used to evaluate several MR biomarkers, including the commonly used non-perfused volume (NPV) measured on contrast-enhanced (CE) T1-weighted (T1w) MR imaging and other biomarkers such as T2w imaging, thermal dose, and diffusion imaging \cite{hectors2014multiparametric,plata2015feasibility,mannelli2009assessment,haider2008dynamic,wu2006registration}.
Although the current spatially non-specific qualitative analysis and indirect metrics demonstrate a strong correlation between some MR biomarkers and histology, the conclusive spatial accuracy cannot be determined with these methods.
For example, conflicting studies show that the NPV biomarker assessed immediately ($\sim$ 1 hour) after treatment both over- and under estimates histology lesion size \cite{futureneedsHectors2016, payne2013vivo}; however, NPV has increased correlation with histology lesion size several days post treatment \cite{futureneedsHectors2016, wijlemans2013evolution}. 
The spatial relationship between \textit{in vivo} MR and histology must be restored to establish the predictive accuracy of \textit{in vivo} MR biomarkers against histology.

Several studies have developed MR to histology registration methods to restore the spatial relationship between \textit{in vivo} diagnostic or treatment evaluation MR images and 2D histology imaging \cite{li2017co,pichat2018survey,losnegaard2018intensity,iglesias2018joint,orczyk2013preliminary,dickinson2013image,goubran2015registration, rusu2020registration}. 
With accurate registration, \textit{in vivo} MR biomarkers can be directly compared to histology using a variety of spatial metrics. 
Sufficient registration accuracy depends on the resolution of the biomarker being evaluated.
Current \textit{in vivo} MR biomarkers are typically acquired with an in-plane resolution of 1-2 mm, for which the ideal registration accuracy is less than one millimeter for accurate evaluation \cite{futureneedsHectors2016, schmitz2010precise}.
A registration technique with this level of accuracy allows development of more accurate, novel biomarkers. 
However, previous registration methods rely on features that correlate directly between \textit{in vivo} MR images and histology that are independent of the diagnostic or treatment features being evaluated \cite{goubran2015registration}.
These independent features can be used to facilitate registration without influencing the final comparison between MR treatment features and the corresponding histology response. 
For example, registration methods for prostate applications excise the entire organ and use whole-mount histology to allow a correlation between the organ boundary on both \textit{in vivo} MR and histology \cite{li2017co,losnegaard2018intensity}.
Many registration methods additionally rely on correlating anatomical landmarks to support the registration \cite{pichat2018survey}.
However, for MRgFUS ablation applications in liver, breast, or kidney, features independent from the ablation treatment features are typically not available for input to registration algorithms. 
This lack of sufficient correlating features between \textit{in vivo} MR and histology independent of the treatment features prohibits the use of previously developed registration techniques. 
Potential treatment dependent features are visible on both \textit{in vivo} MR and histology, but the use of these features to drive registration biases the final result of biomarker evaluation. 
The need is critical for a new registration method that uses intermediate steps in lieu of feature correlation to accurately register \textit{in vivo} MR and histology to evaluate MRgFUS treatment biomarkers. 

In this paper, we present a novel, multi-step MR to histology registration workflow that corrects all deformations without assuming any direct feature correlation between \textit{in vivo} MR and histology, but rather between three intermediate stages. 
We estimate the deformation from every tissue processing step and compose the deformations to generate a full 3D map between any histology image and \textit{in vivo} MR biomarkers.
We densely sample the histology space and map all histology annotations to \textit{in vivo} MR to form a volumetric histological annotation of tissue necrosis in the \textit{in vivo} MR space. 
This novel registration workflow facilitates an evaluation of MR biomarkers that was not possible before, using spatial metrics such as precision, recall, DICE coefficient, and Hausdorff distance \cite{taha2015metrics}.
We demonstrate the capabilities of this registration pipeline by evaluating the NPV biomarker following MRgFUS ablation of a VX2 rabbit tumor model at two time points: 1) acute NPV ($\sim$ 1 hour after ablation) and 2) post-NPV (3-5 days after treatment).
We compare these spatially quantified results to the findings of qualitative and indirect findings in the literature.
The presented registration method will facilitate accurately validating a wide range of \textit{in vivo} MR biomarkers against histology.
Validating \textit{in vivo} MR biomarkers will increase their clinical viability for MRgFUS treatment assessment and facilitate minimally invasive treatments becoming feasible alternatives to surgical intervention for cancer treatments. 

\begin{figure}[ht]
	\centering
	\includegraphics[width=0.45\textwidth]{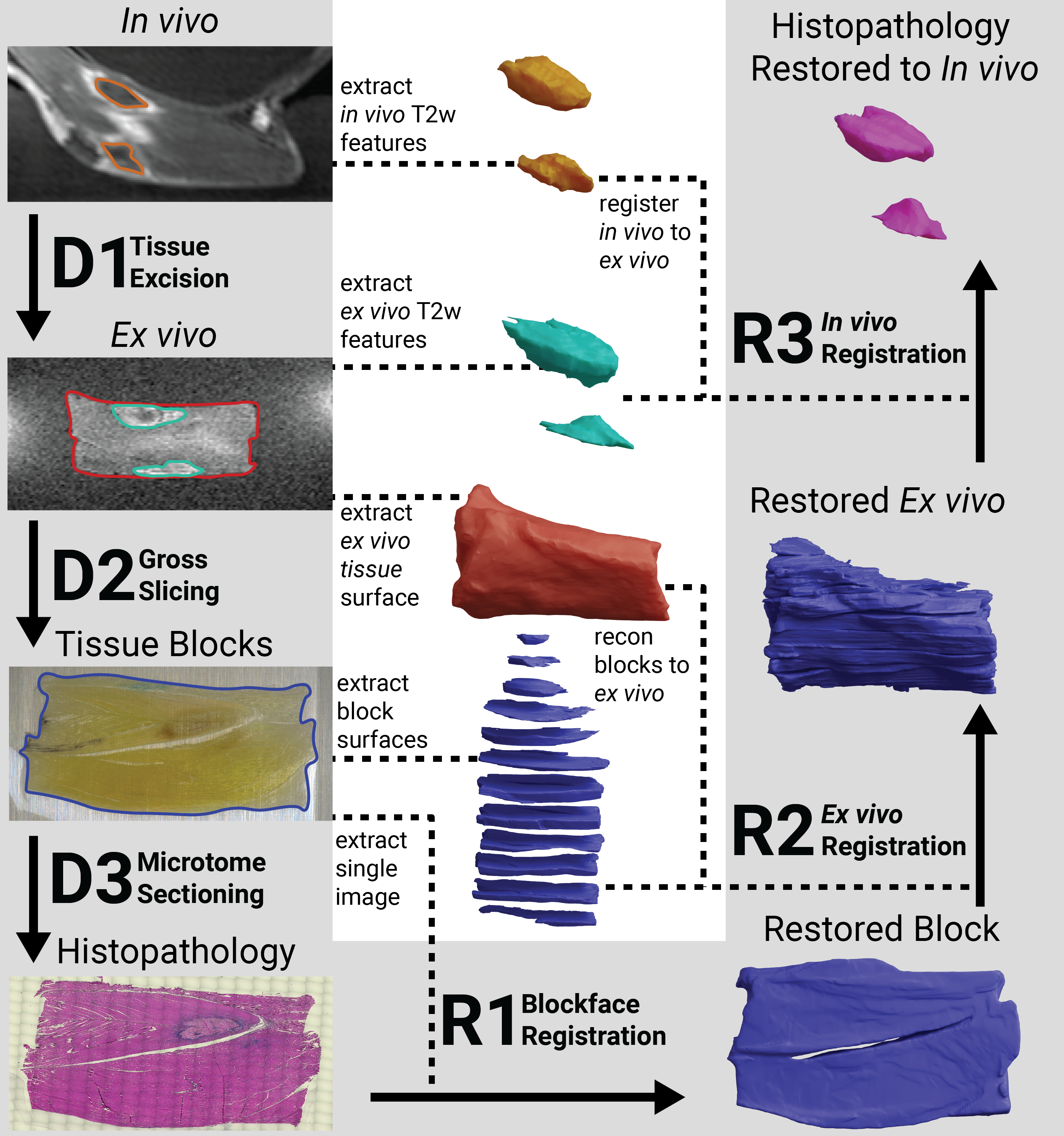}
	\caption{Flow chart of the end-to-end workflow required to register \textit{in vivo} MR to histology. The steps for destructive histopathology pipeline are indicated by D1, D2, and D3, and the steps for restoring registration pipeline are indicated by R1, R2, and R3. Dashed lines indicate information extracted from digital imaging during the destructive histopathology pipeline for use in the restoring steps.\vspace{-6mm}}
	\label{fig:flowchart}
\end{figure}

\vspace{-1em}
\section{Overview of Approach}
\begin{figure}[t]
\centering
    \begin{tikzpicture}\draw (0, 0) node[inner sep=0]
    {\includegraphics[]{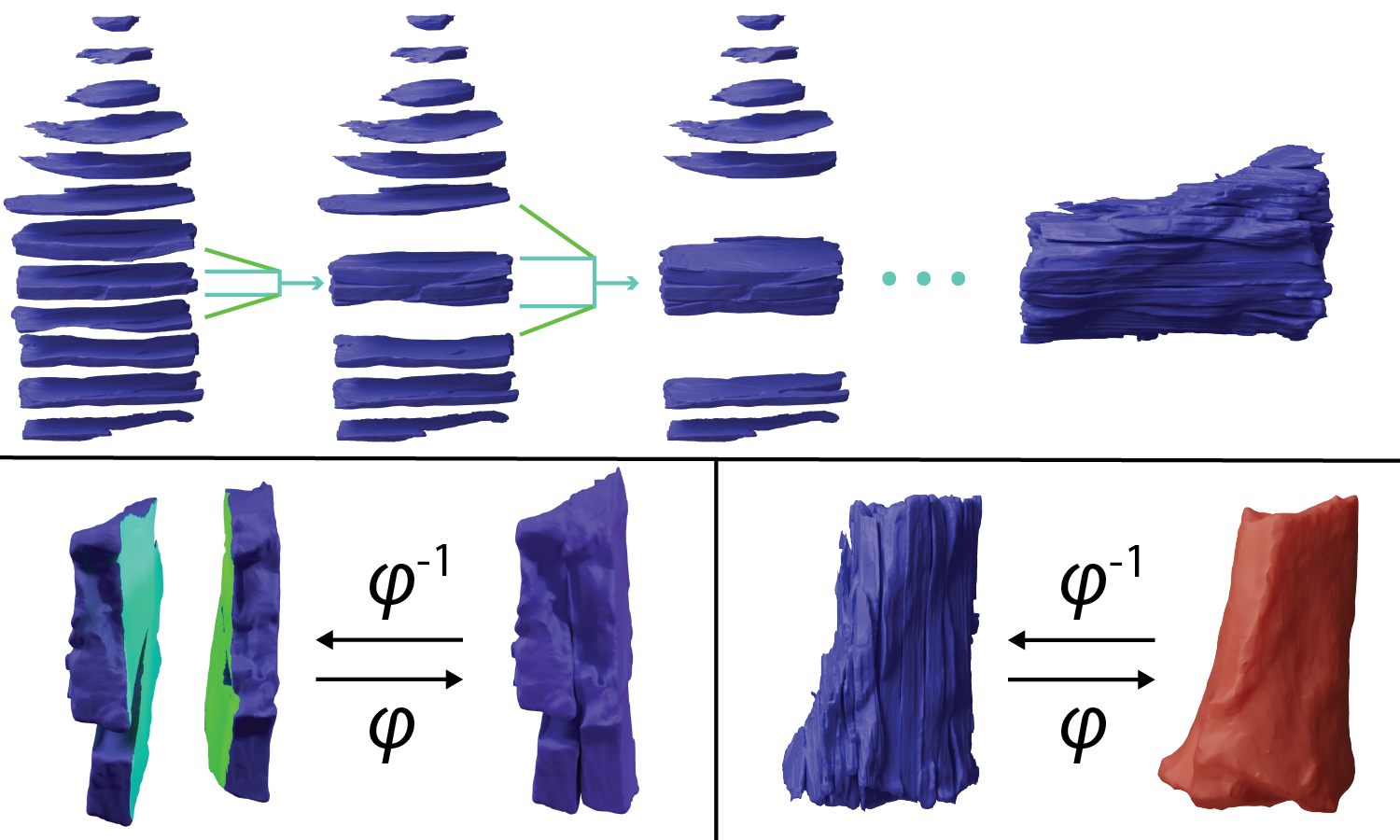}};
    \draw ((-6.2, 3.5) node[text=black]{(a)};
    \draw ((-6.2,-0.8) node[text=black]{(b)};
    \draw ((0.45,-0.8) node[text=black]{(c)};
    \end{tikzpicture}
\caption{Depiction of the block reconstruction process. The sequential block reconstruction is shown in (a). Two blocks are registered together using corresponding faces as in (b), where the light blue face is the target face and does not move, and the light green face is the moving face. This registration results in a transformation $\varphi$ for the moving block. The reconstructed block is registered to the red \textit{ex vivo} surface in (c) to generate a transformation $\varphi$ between the reconstructed blocks and \textit{ex vivo} MR imaging. \vspace{-4mm}}
\label{fig:block_recon}
\end{figure}
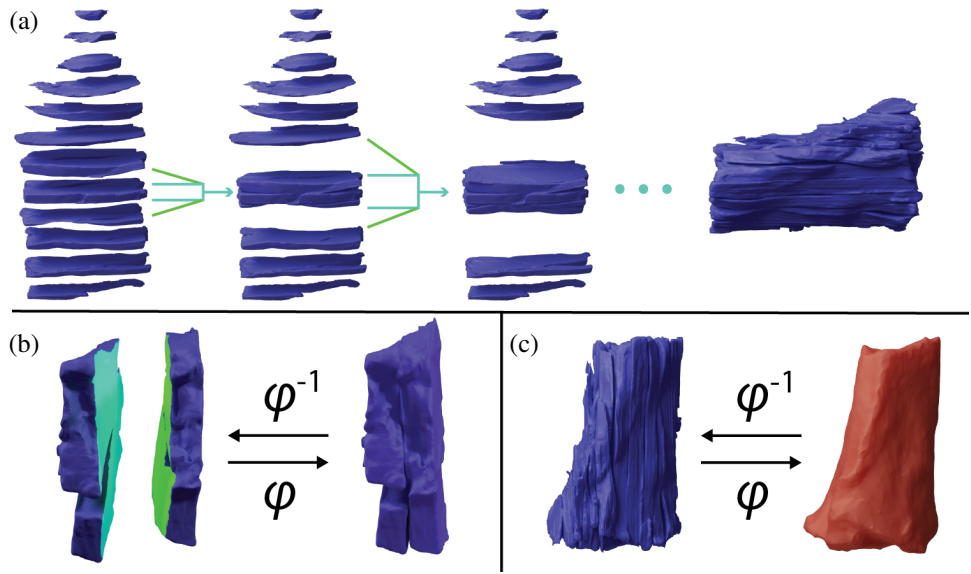

The presented registration approach utilizes intermediate imaging to comprehensively correct for deformations introduced throughout the tissue processing workflow. 
Intermediate imaging provides digital representations of the tissue both before and after specific deformations have been introduced during processing, eliminating the need for correlating features between \textit{in vivo} MR and histology to drive the registration, as has been done in previous methods. 
A depiction of the general workflow is shown in \figureautorefname~\ref{fig:flowchart}. 
The pipeline developed and described in this work combines previously developed registration strategies and contributes novel registration methods \cite{pichat2018survey}.
The tissue processing and registration pipelines were evaluated on four subjects of a large animal tumor model described in \sectionautorefname~\ref{sec:model}.

The (D)estructive histopathology pipeline seen in \figureautorefname~\ref{fig:flowchart} consists of three main steps that introduce deformation into the tissue: tissue excision (D1), gross slicing (D2), and microtome sectioning (D3).
Surgical excision and formalin fixation of the treated tissue introduces orientation uncertainty and non-linear deformation into the tissue sample.
The excised tissue is typically too large to be processed as a whole, so the sample is grossly sliced into smaller tissue blocks, usually $\sim$3-5mm. 
The gross slicing step introduces non-linear deformations to each tissue block and destroys the spatial relationship between each block and the original volume.
In typical histology processing, after paraffin wax embedding, each block face is trimmed using a microtome until a full section of tissue is exposed ('facing the block'), and then sections are collected for histology and the remaining tissue is set aside \cite{pichat2018survey}. 
However, the presented method requires microtome sectioning through the entire tissue block, collecting sections for histology throughout the block. 
Microtome sectioning and subsequent staining and mounting cause non-linear deformations from tissue shearing, tearing, and stretching.
The end product of the destructive histopathology pipeline is a series of stained 2D microscopic images that must be re-aligned with \textit{in vivo} MR.

The (R)estoring registration pipeline independently estimates the deformation between each of three main destructive histopathology steps via blockface registration (R1), \textit{ex vivo} registration (R2), and \textit{in vivo} registration (R3).
Blockface registration (R1) estimates the 2D deformations from microtome sectioning, staining, and mounting.
After blockface registration, each 2D histology image must be correlated with the 3D tissue volume before gross slicing. 
In this step, most other registration methods depend on 2D slice correlation between MR (\textit{in vivo} or \textit{ex vivo}) and histology to restore the relationship between the original tissue volume and histology \cite{pichat2018survey}. 
However, our novel \textit{ex vivo} registration (R2) step uses 3D models from \textit{ex vivo} MR and blockface imaging to embed each histology section into the original 3D tissue volume without assuming slice-to-slice correspondence or correlation between MR and histology features.
A general overview of this reconstruction step is shown in \figureautorefname~\ref{fig:block_recon}.
We emulate the 2D slice correspondence assumption in our \textit{ex vivo} registration step and compare the error of the presented method against this assumption to show the benefit of the 3D model reconstruction.
Finally, \textit{in vivo} registration (R3) estimates deformations from tissue excision by registering 3D models of corresponding features from \textit{ex vivo} and \textit{in vivo} MR images. 
The deformations between any two steps is represented by a diffeomorphism, and these deformations can be composed to generate a single diffeomorphism between multiple stages of imaging. 
A single diffeomorphism from histology to \textit{in vivo} MR inserts each histology section into the \textit{in vivo} MR space and enables the generation of fully registered, volumetric histological necrosis annotations in the \textit{in vivo} MR space.

The necessary accuracy of registration is dependent on the native resolution of the evaluation space, which is \textit{in vivo} MR imaging. 
Ideally, the registration accuracy should be at or below the resolution of the acquired MR biomarkers being evaluated. 
Common clinical biomarkers are acquired with an in-plane resolution of 1-2 mm, making 1 mm of total error an ideal constraint for the total registration error \cite{futureneedsHectors2016,schmitz2010precise}. 
The accuracy of the presented registration pipeline was evaluated at every stage using the Euclidean distance between corresponding landmarks following registration, known as the target registration error (TRE).
These landmarks are only between intermediate stages, not directly between \textit{in vivo} MR and histology, and are not used to drive registration.
For example, features to evaluate R3 between \textit{in vivo} T2w MR and \textit{ex vivo} T2w MR can be identified because they are the same imaging modality. 
Additionally, landmarks, such as the the edge of the tissue, can be selected between histology sections and blockface images to evaluate R1. 
However, the edge of the tissue is not identifiable on \textit{in vivo} MR (because the tissue was trimmed) and the appearance of the T2w MR features in histology may not be known.  
Unlike other applications, selecting landmarks that correlate between \textit{in vivo} MR and histology is not possible in this application, so we evaluate the individual stages of registration.  
The total accuracy of the pipeline is estimated by summing the errors from each stage to obtain the cumulative error from registration.
To determine the potential impact of the 2D slice projection on the registration accuracy, the landmarks from \textit{ex vivo} registration (R2) are deformed using the proposed registration method and the 2D slice assumption method independently and compared.
The acute NPV and post-NPV biomarkers are spatially evaluated against the volumetric histological annotation of necrosis resulting from registration using spatially specific metrics, including the precision, recall, DICE coefficient, and Hausdorff distance.

\newlength{\volcompwidth}
\setlength{\volcompwidth}{5cm}
\makeatletter
\newcommand{\definetrim}[2]{%
  \define@key{Gin}{#1}[]{\setkeys{Gin}{trim=#2,clip}}%
}
\makeatother

\definetrim{vol_trim}{0.4in 0.4in 0.3in 0.6in}
\begin{figure}[ht]
\centering
\begin{minipage}{\textwidth}%
    \centering%
    \begin{subfigure}[b]{0.95\textwidth}%
    \centering%
        \begin{tikzpicture}\draw (-5.3, 0) node[inner sep=0]
        {\includegraphics[width=10cm, trim={0.4in 0.4in 0.3in 0.6in}, clip]{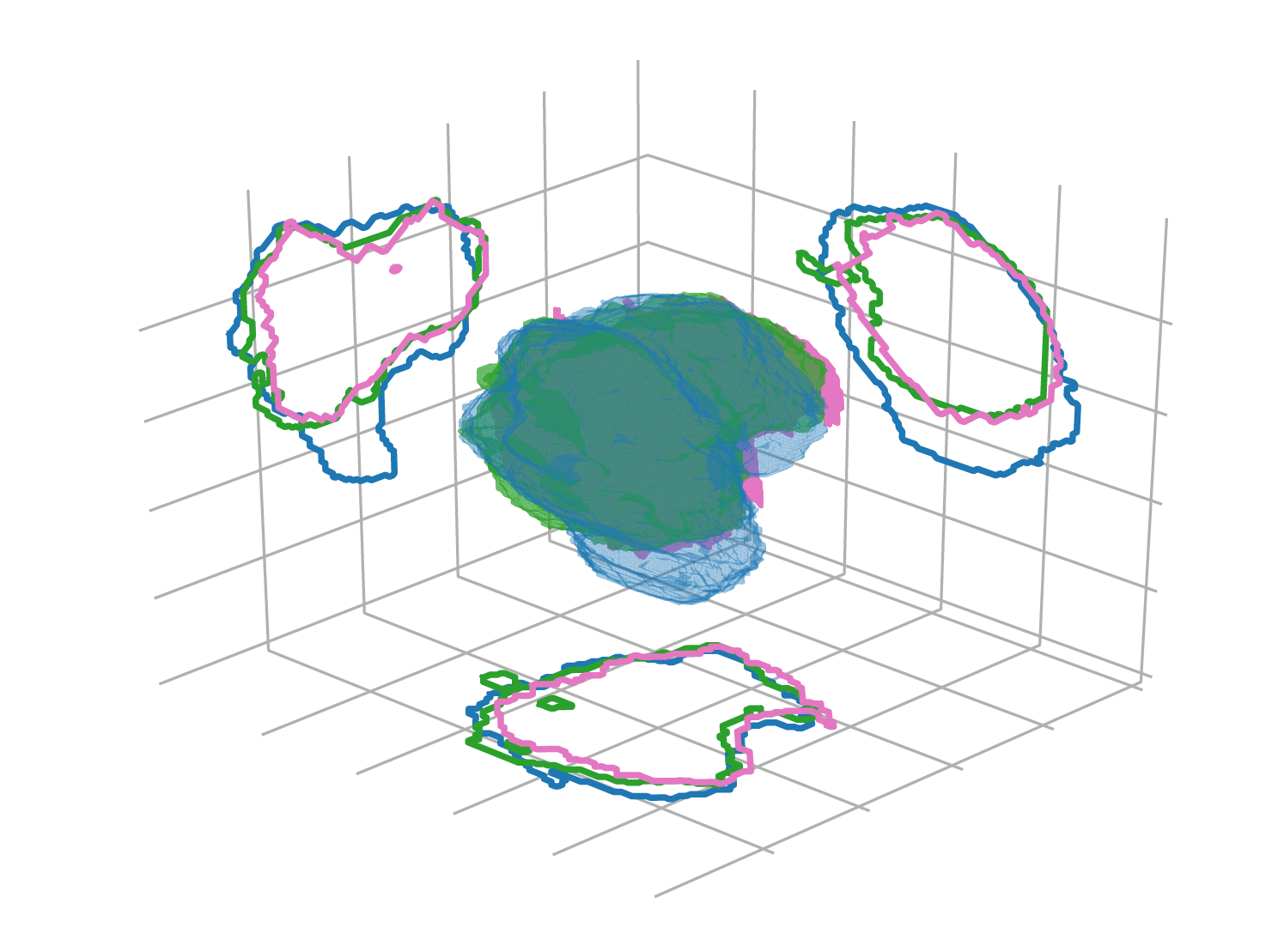}};
        \draw ((-9.5,-2.7) node[text=black]{(a)};
        \draw ((1.5,-2.0) node[text=black]{(c)};
        \draw ((1.5,0.5) node[text=black]{(b)};
        \draw (3.9, -1.3) node[inner sep=0]
        {\includegraphics[width=4.7cm, angle=7, trim={0.6in 0.6in 0.7in 0.8in}, clip]{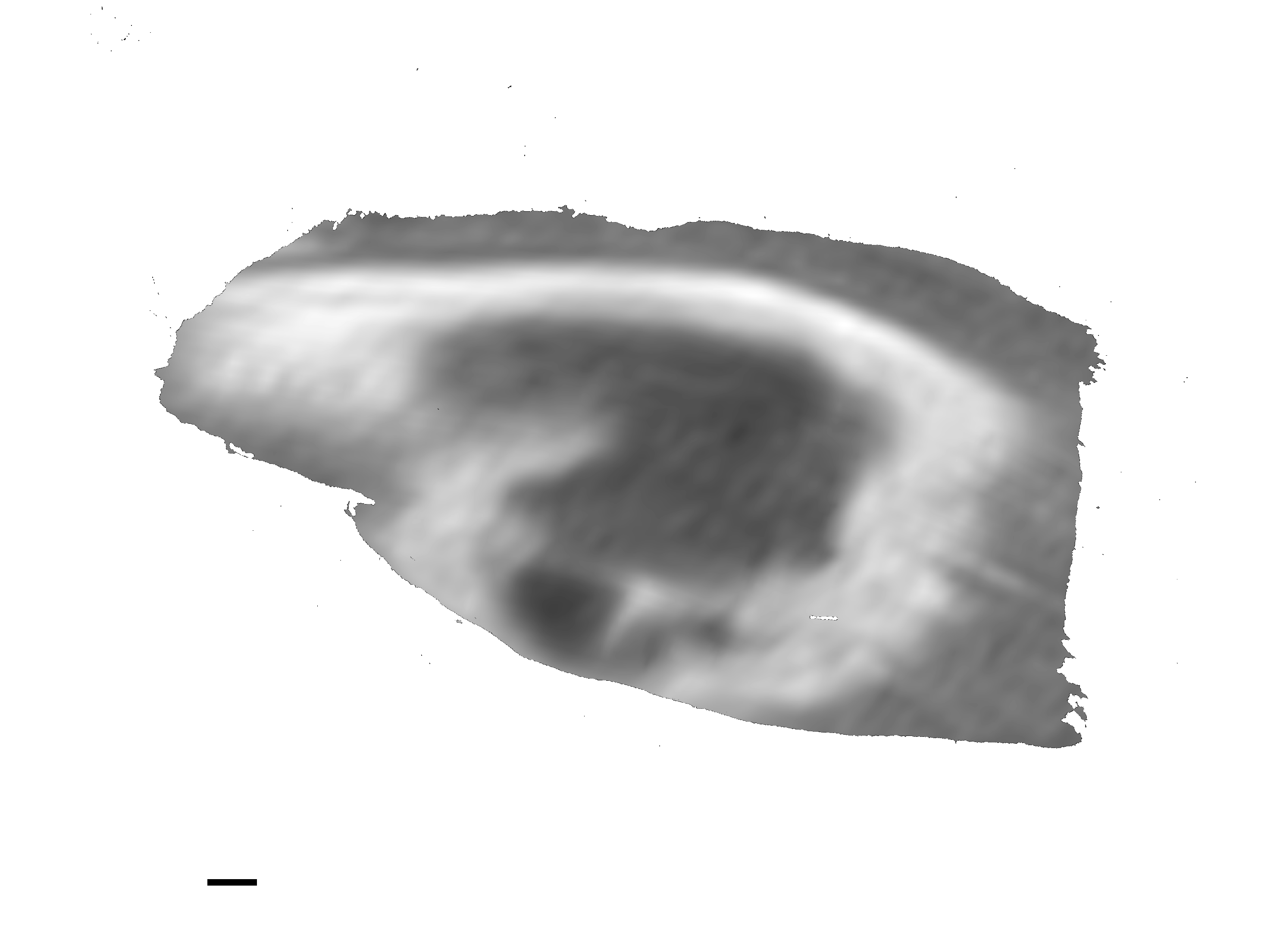}};
        \draw (3.9, -1.3) node[inner sep=0]
        {\includegraphics[width=4.7cm, angle=7, trim={0.6in 0.6in 0.7in 0.8in}, clip]{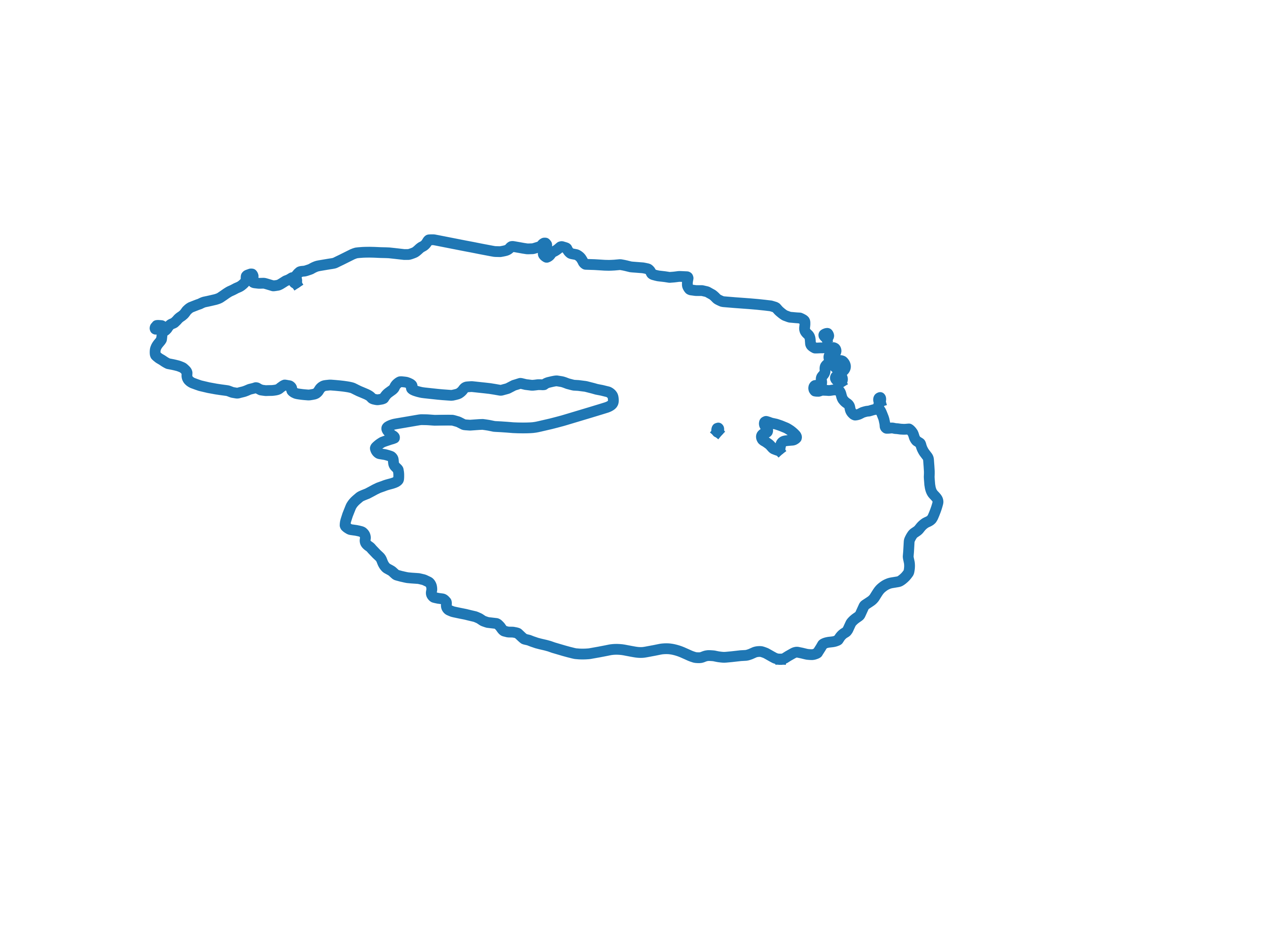}};
        \draw (3.9, -1.3) node[inner sep=0]
        {\includegraphics[width=4.7cm, angle=7, trim={0.6in 0.6in 0.7in 0.8in}, clip]{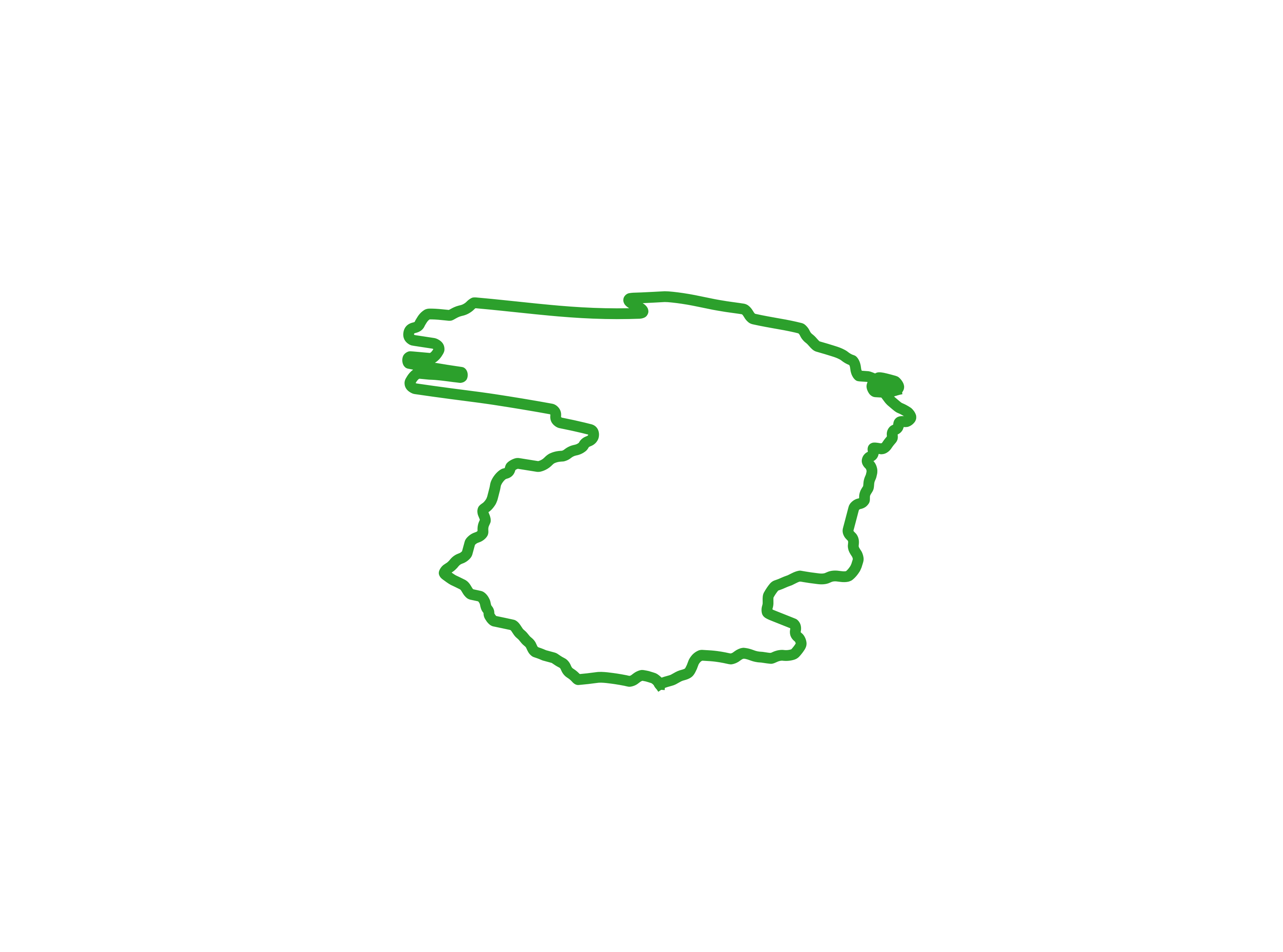}};
        \draw (3.9, -1.3) node[inner sep=0]
        {\includegraphics[width=4.7cm, angle=7, trim={0.6in 0.6in 0.7in 0.8in}, clip]{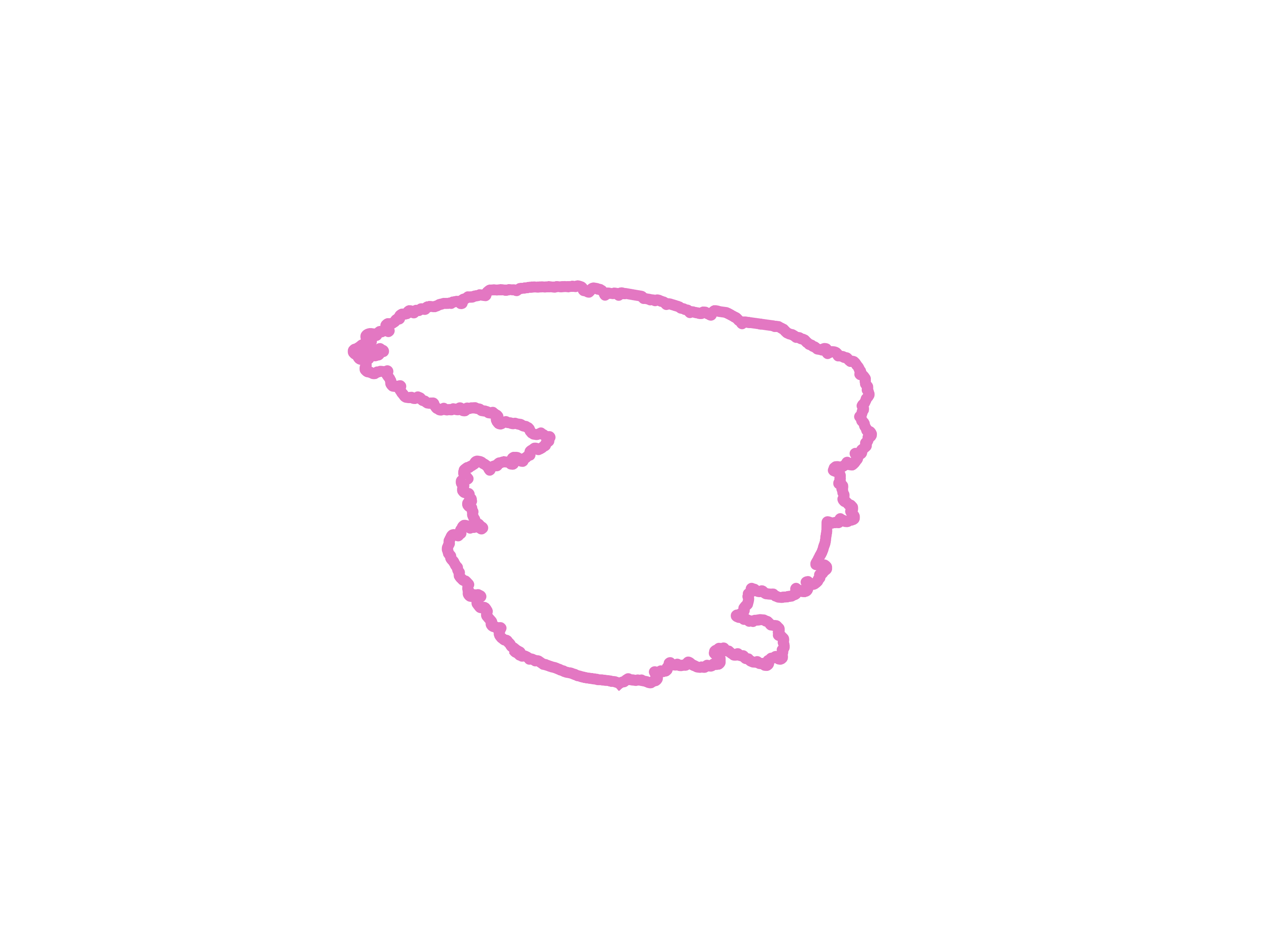}};
        \draw (3.9, 1.3) node[inner sep=0]
        {\includegraphics[width=4.7cm, angle=7, trim={0.6in 0.6in 0.7in 0.8in}, clip]{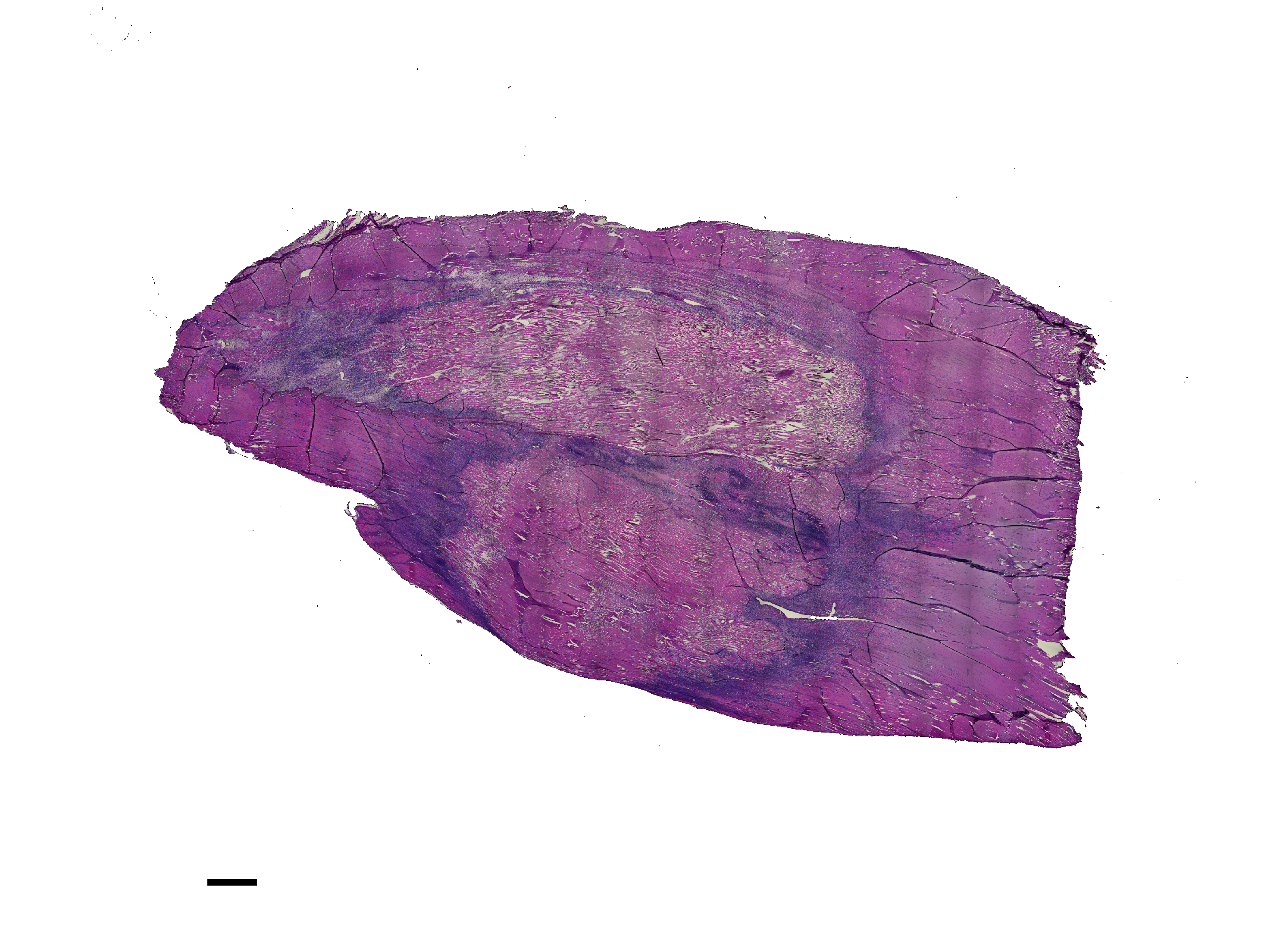}};
        \draw (3.9, 1.3) node[inner sep=0]
        {\includegraphics[width=4.7cm, angle=7, trim={0.6in 0.6in 0.7in 0.8in}, clip]{Figures/npv_volume_comp/18_062_d0_npv.png}};
        \draw (3.9, 1.3) node[inner sep=0]
        {\includegraphics[width=4.7cm, angle=7, trim={0.6in 0.6in 0.7in 0.8in}, clip]{Figures/npv_volume_comp/18_062_ce_npv.png}};
        \draw (3.9, 1.3) node[inner sep=0]
        {\includegraphics[width=4.7cm, angle=7, trim={0.6in 0.6in 0.7in 0.8in}, clip]{Figures/npv_volume_comp/18_062_hist_npv.png}};
        \end{tikzpicture}
    \end{subfigure}
    \\\vspace{2mm}\hrule\vspace{2mm}
    \hfill
    \begin{subfigure}[b]{0.32\textwidth}
    \centering
        \begin{tikzpicture}\draw (0, 0) node[inner sep=0]
        {\includegraphics[width=\volcompwidth, vol_trim, clip]{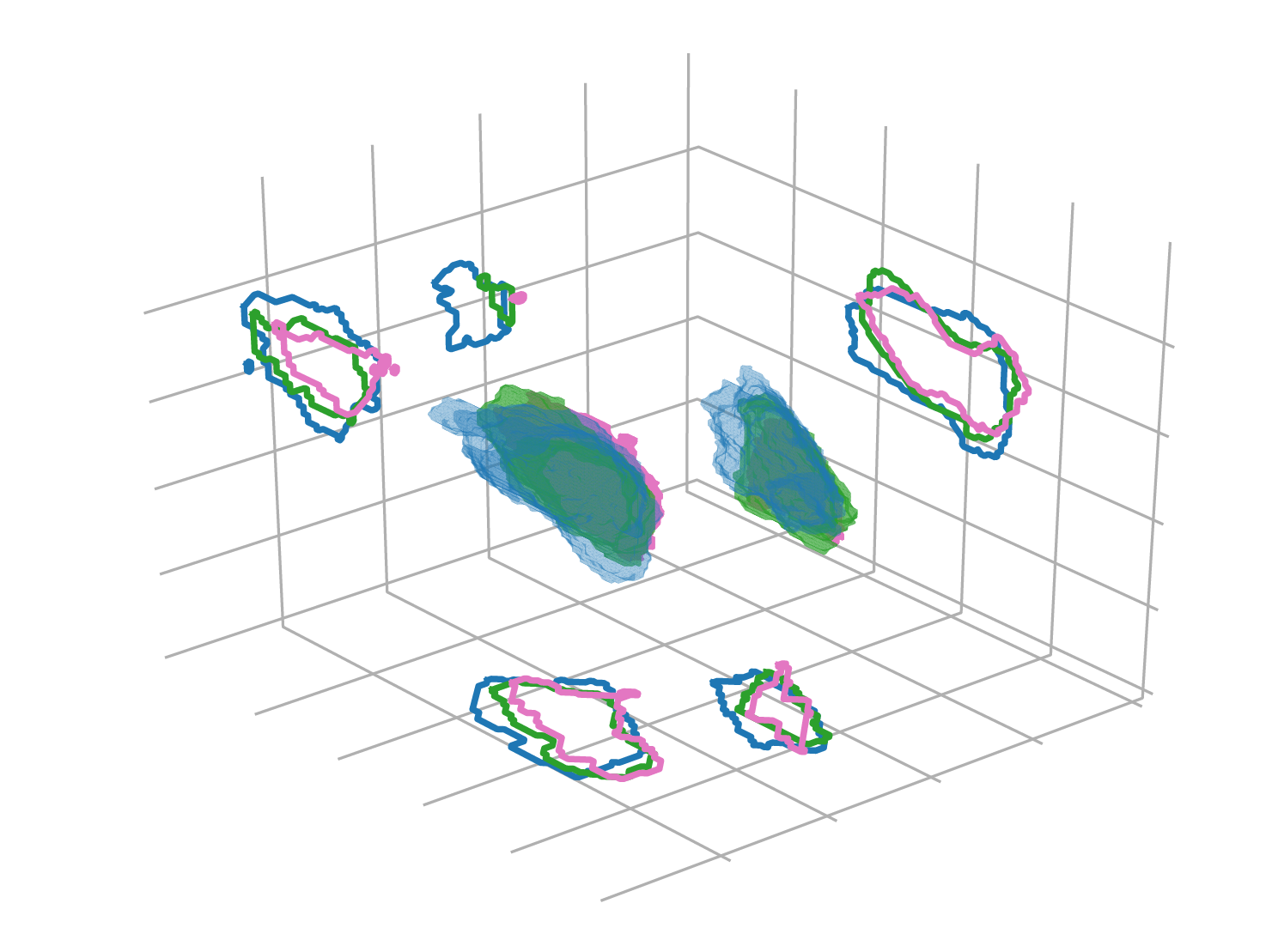}};
        \draw ((-1.8,-1.5) node[text=black]{(d)};
        \end{tikzpicture}
    \end{subfigure}%
    \hfill%
    \begin{subfigure}[b]{0.32\textwidth}
    \centering
        \begin{tikzpicture} \draw (0, 0) node[inner sep=0]
        {\includegraphics[width=\volcompwidth, vol_trim, clip]{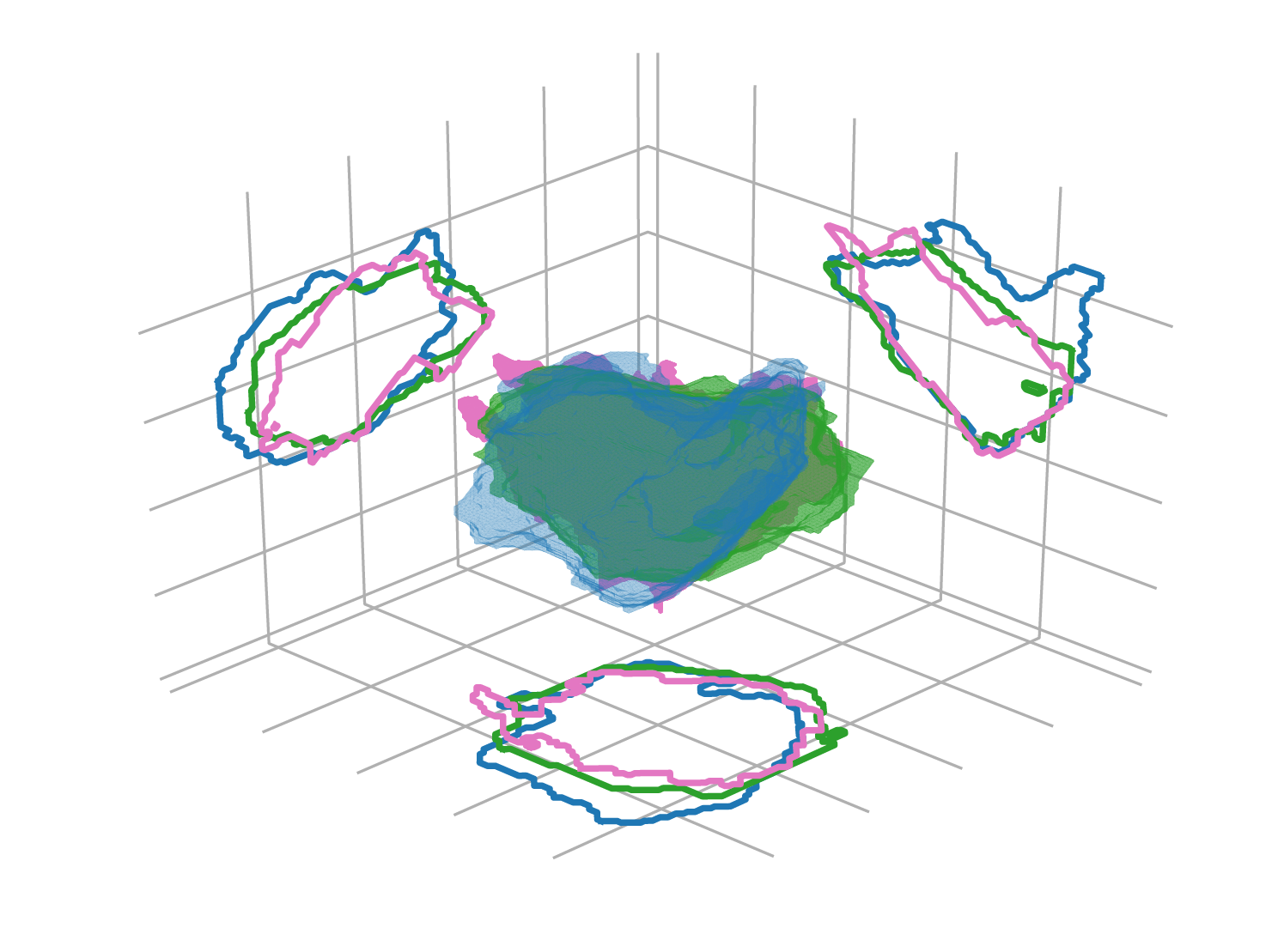}};
        \draw ((-1.8,-1.5) node[text=black]{(e)};
        \end{tikzpicture}
    \end{subfigure}%
    \hfill%
    \begin{subfigure}[b]{0.32\textwidth}
    \centering
        \begin{tikzpicture}\draw (0, 0) node[inner sep=0] 
        {\includegraphics[width=\volcompwidth, vol_trim, clip]{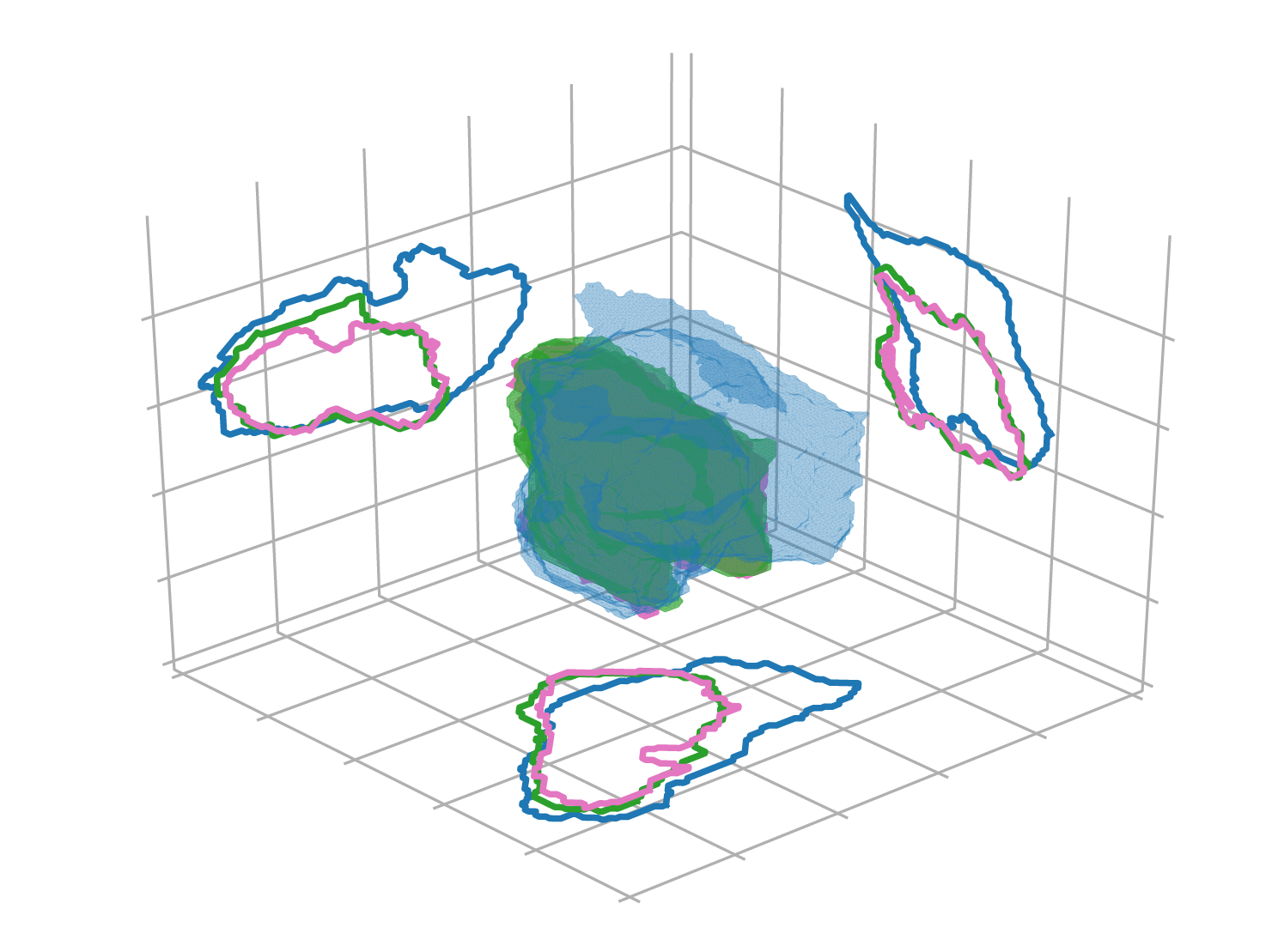}};
        \draw ((-1.8,-1.5) node[text=black]{(f)};
        \end{tikzpicture}
    \end{subfigure}\hfill%
    \end{minipage}
    \vspace{-4pt}
    \caption{MR biomarker to histology registration with accurate spatial comparisons of the acute NPV (blue), post-NPV (green), and histology necrosis (pink) labels. (a, d-f) show volumetric overlays of each of the labels in the \textit{in vivo} MR space for all four subjects. The contours on the axis walls are projections for each respective volume from the slice with the largest acute NPV area in each dimension. The space between each grid line is 5 mm. (b) Label contours from the subject shown in (a) sampled into the native histology space. (c) Follow-up contrast-enhanced MR image resampled onto the corresponding histology slice in (b) with the each of the labels overlaid, demonstrating that the estimated transformations are invertible. \vspace{-4mm} }
    \label{fig:volume_comp}
\end{figure}
\vspace{-1em}
\section{Results}

The goal of the registration pipeline was to provide a 3D volumetric label of histological necrosis registered to the \textit{in vivo} MR imaging space. 
\figureautorefname~\ref{fig:volume_comp} includes a volumetric comparison between the acute NPV (blue) and  post-NPV (green) MR biomarkers and the histology necrosis (pink) for each of the four subjects analyzed. 
The histology volume represents the label of tissue necrosis against which the acute and post NPV MR biomarkers were independently compared to determine their accuracy. 
The spatial metrics calculated from these volumes are presented in \tableautorefname~\ref{tab:spatial_metrics}.
The TRE accuracy for each stage and specific stages under different assumptions is displayed in \figureautorefname~\ref{fig:stage_tre}.

\begin{table}[b]
\definecolor{columnGray}{gray}{0.90}
  \centering
    \caption{The total volume and different spatial evaluation metrics for each label, the acute NPV and post-NPV, compared against the histology volume for each subject analyzed.}
    \label{tab:spatial_metrics}
    \begin{filecontents*}{Figures/spatial_comp/metrics.txt}
sub1 sub2 sub3 sub4 sub5 sub6 sub7 sub8 sub9 sub10 sub11 sub12
1 1317.35 915.74 675.59 0.39 0.62 0.76 0.85 0.52 0.72 7.61 3.17
2 2039.13 1688.73 1281.84 0.49 0.69 0.78 0.91 0.60 0.78 6.09 3.31
3 3859.13 2627.20 2391.17 0.49 0.86 0.80 0.94 0.61 0.90 7.51 3.05
4 5542.83 3969.91 3477.31 0.58 0.81 0.92 0.92 0.71 0.86 12.14 3.07
\end{filecontents*}
\pgfplotstabletypeset[
    every odd column/.style={column type/.add={>{\columncolor[gray]{.8}}}{}},
      multicolumn names, 
      display columns/0/.style={column name=\#},  
      display columns/1/.style={column name=\shortstack{\footnotesize Acute\\ \footnotesize NPV  \\  \footnotesize Vol. ($mm^3$)}, fixed zerofill},
      display columns/2/.style={column name=\shortstack{\footnotesize Post \\  \footnotesize NPV \\  \footnotesize Vol. ($mm^3$)}, fixed zerofill},
      display columns/3/.style={column name=\shortstack{\footnotesize Histology \\ \footnotesize  Vol. ($mm^3$)}, fixed zerofill},
      display columns/4/.style={column name=\shortstack{\footnotesize Acute \\  \footnotesize NPV \\ \footnotesize Precision}, fixed zerofill},
      display columns/5/.style={column name=\shortstack{\footnotesize Post \\  \footnotesize NPV \\ \footnotesize Precision}, fixed zerofill},
      display columns/6/.style={column name=\shortstack{\footnotesize Acute \\  \footnotesize NPV \\ \footnotesize Recall}, fixed zerofill},
      display columns/7/.style={column name=\shortstack{\footnotesize Post \\  \footnotesize NPV \\ \footnotesize Recall}, fixed zerofill},
      display columns/8/.style={column name=\shortstack{\footnotesize Acute \\  \footnotesize NPV \\ \footnotesize DICE}, fixed zerofill},
      display columns/9/.style={column name=\shortstack{\footnotesize Post \\  \footnotesize NPV \\ \footnotesize DICE}, fixed zerofill},
      display columns/10/.style={column name=\shortstack{\footnotesize Acute NPV \\ \footnotesize Hausdorff ($mm$)}, fixed zerofill},
      display columns/11/.style={column name=\shortstack{\footnotesize Post NPV \\ \footnotesize Hausdorff ($mm$)}, fixed zerofill},
      every head row/.style={before row={\toprule}, 
		after row={\midrule}},
		every last row/.style={after row=\bottomrule}, 
    ]{Figures/spatial_comp/metrics.txt} 
    \vspace{-8pt}
\end{table}
\subsection{Registration Pipeline Accuracy}

\makeatletter
\newcommand\resetstackedplotsfive{
\makeatletter
\pgfplots@stacked@isfirstplottrue
\makeatother
\addplot [forget plot,draw=none] coordinates{(1,0) (2,0) (3,0) (4,0)};
}

\def\datafile{data.csv}

\definecolor{cat1_color}{RGB}{0,48,143}%
\definecolor{cat2_color}{RGB}{25,116,210}%
\definecolor{cat3_color}{RGB}{145,223,255}%
\definecolor{cat4_color}{RGB}{178,255,255}%
\definecolor{cat5_color}{RGB}{223,41,32}%
\definecolor{cat6_color}{RGB}{255,162,186}%
\begin{figure}[tb]
\centering
\begin{minipage}{0.43\textwidth}
    \centering
\begin{tikzpicture}
\tikzstyle{every node}=[font=\small]
\begin{axis}[
    width=0.9\columnwidth,
    scale only axis=true,
    height=5cm,
    xtick={1,...,4},
    xticklabels={1,2,3,4},
    enlarge x limits=0.2,
    ymajorgrids,
    major tick length = 0,
    ylabel = TRE (mm),
    ymax=1.7,
    xlabel={Subject Number},
    ybar=1pt,
    title={Proposed Pipeline Per Stage Error},
    legend style={at={(0.5,0.97)},anchor=north, legend cell align=left, align=left, draw=black}, legend columns=2
    ]
\addplot[
    fill=cat1_color,
    draw=black,
    point meta=y,
    every node near coord/.style={inner ysep=5pt},
    error bars/.cd,
        y dir=both,
        y explicit
] 
table [x=num, y error=r1ds, y=r1dm, col sep=comma] {Figures/landmark_comp/stats.csv};
\addlegendentry{R1 Error \hspace{5mm}} 
\addplot[
    fill=cat2_color,
    draw=black,
    point meta=y,
    every node near coord/.style={inner ysep=5pt},
    error bars/.cd,
        y dir=both,
        y explicit
] 
table [x=num, y error=r2ds, y=r2dm, col sep=comma] {Figures/landmark_comp/stats.csv};
\addlegendentry{R2 Error}
\addplot[
    fill=cat3_color,
    draw=black,
    point meta=y,
    every node near coord/.style={inner ysep=5pt},
    error bars/.cd,
        y dir=both,
        y explicit
] 
table [x=num, y error=r3ds, y=r3dm, col sep=comma] {Figures/landmark_comp/stats.csv};
\addlegendentry{R3 Error}
\addplot[
    fill=cat4_color,
    draw=black,
    point meta=y,
    every node near coord/.style={inner ysep=5pt},
    error bars/.cd,
        y dir=both,
        y explicit
] 
table [x=num, y error=tdm, y=tds, col sep=comma] {Figures/landmark_comp/stats.csv};
\addlegendentry{Cum. Error}
\draw ({rel axis cs:0,0}|-{axis cs:0,0}) -- ({rel axis cs:1,0}|-{axis cs:0,0});
\end{axis}
\end{tikzpicture}
\end{minipage}
\begin{minipage}{0.49\textwidth}
    \centering
\begin{tikzpicture}
\tikzstyle{every node}=[font=\small]
\begin{axis}[set layers,
    width=.9\columnwidth,
    major tick length = 0,
    ylabel = TRE (mm),
    xlabel={Subject Number},
    ymajorgrids,
    height=6.65cm,
    ymin=-.1,
    ymax=1.7,
    xticklabels={1,2,3,4},
    ybar stacked,
    xtick={1,2,3,4},
    title={2D Slice Projection Comparison},
    legend style={at={(0.5,0.97)},anchor=north, legend cell align=left, align=left, draw=black},
    legend columns=2
]
\addplot
    [forget plot, fill=cat2_color, discard if={Name}{LR}]
    table [x=X, y=Activation] {\datafile};
\addplot
    [
    on layer=axis foreground,
    fill=cat2_color, discard if={Name}{LR},
    error bars/.cd, y dir=both, y explicit,
    ]
    table [x=X, y=Inclusion, y error plus=Max, y error minus=Min] {\datafile};
\addlegendentry{R2 Error \hspace{1mm}}
\resetstackedplotsfive
\addplot
    [black, fill=cat5_color, discard if={Name}{CR}]
    table [x=X, y=Activation] {\datafile};
\addplot
    [on layer=axis foreground,
    fill=cat6_color, discard if={Name}{CR}, error bars/.cd, y dir=both, y explicit,]
    table [x=X, y=Inclusion, y error plus=Max, y error minus=Min] {\datafile};
\addplot
    [
    on layer=axis foreground,
    fill=cat2_color, discard if={Name}{LR},
    error bars/.cd, y dir=both, y explicit
    ]
    table [x=X, y=Inclusion] {\datafile};
\addlegendentry{2D Slice}
\end{axis}
\end{tikzpicture}
\end{minipage}
\vspace{-10pt}
\caption{Target registration error (TRE) computed on a subject-specific basis. (left) Subject-specific TREs for the three stages of the registration pipeline: blockface (R1), \textit{ex vivo} (R2), and \textit{in vivo} (R3) registration. The cumulative error is the summation of the mean of three stages for each subject. (right) Subject-specific TREs from \textit{ex vivo} registration (R2) for the presented method versus the 2D slice correlation method. The dark region of the 2D slice bar shows the portion of the total error resulting only from the Z dimension. \vspace{-5mm}}
\label{fig:stage_tre}
\end{figure}
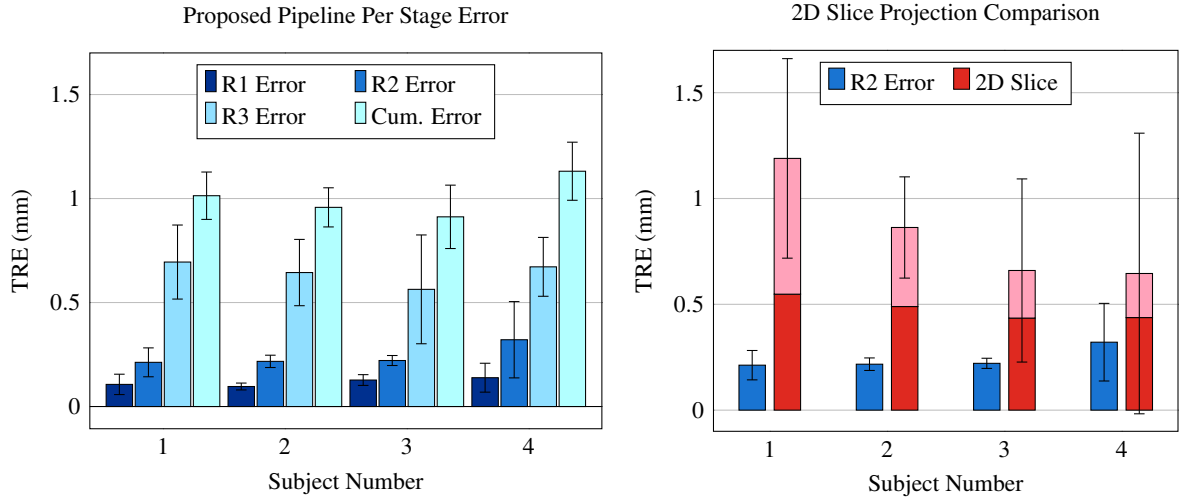 

The accuracy of each stage of the registration pipeline was independently evaluated using landmarks; the R1 landmarks were different from the R2 landmarks, which were both different from the R3 landmarks.
TREs for the different stages of the registration pipeline are plotted in \figureautorefname~\ref{fig:stage_tre} for the four subjects analyzed.
The TRE for blockface registration (R1), \textit{ex vivo} registration (R2), and \textit{in vivo} registration (R3) ranged from 0.096 to 0.139 mm (mean $\pm$ s.d. = 0.117 $\pm$ 0.045 mm), 0.21 to 0.32 mm (0.24 $\pm$ 0.10 mm), and 0.56 to 0.69 mm (0.64 $\pm$ 0.19 mm), respectively. 
The sum of the error for the different stages ranged from 0.91 to 1.13 mm (1.00 $\pm$ 0.13 mm) for the four subjects analyzed. 

\subsection{Comparison with 2D Slice Correlation}

The \textit{ex vivo} registration (R2) was run while restricting the Z dimension (perpendicular to the histology images) motion to only translation, enforcing the assumption that the blockface and 2D histology slices correlate with an \textit{ex vivo} MR slice. 
The TRE for the presented \textit{ex vivo} registration method and the 2D slice correlation method can be seen in \figureautorefname~\ref{fig:stage_tre}.
The presented method ranged from 0.21 to 0.32 mm (0.24 $\pm$ 0.10 mm) and significantly (p = 0.025, N=20) outperformed the 2D slice correlation method, which ranged from 0.65 to 1.19 mm (0.84 $\pm$ 0.22 mm).
The 3D error from the presented method even significantly (p=0.015, N=20) outperformed the Z dimension error of the 2D slice correlation method alone, ranging from 0.44 to 0.54 mm (0.48 $\pm$ 0.05 mm). 

\subsection{NPV Biomarker Evaluation}

The total volume of each label and the spatial metrics comparing the acute NPV and the post NPV MR biomarkers against the histology necrosis label are shown in \tableautorefname~\ref{tab:spatial_metrics}.
The post-NPV precision significantly outperformed the acute NPV precision (0.0079, N=4) when compared to the histology necrosis labels, ranging from 0.62 to 0.86 (0.74 $\pm$ 0.09) and 0.39 to 0.58 (0.49 $\pm$ 0.07), respectively. 
The post-NPV recall ranged from 0.76 to 0.94 (0.91 $\pm$ 0.03) but was not significantly different (p=0.079, N=4) from the acute NPV that ranged from 0.76 to 0.92 (0.82 $\pm$ 0.06). 
The DICE similarity coefficient was significantly better (p=0.011, N=4) between post-NPV and histology, which ranged from 0.72 to 0.9 (0.82 $\pm$ 0.07), than between the acute NPV and histology, which ranged from 0.52 to 0.71 (0.61 $\pm$ 0.07).
Finally, the Hausdorff distance for post-NPV ranged from 3.05 to 3.31 mm (3.15 $\pm$ 0.10 mm) and was significantly smaller (p = 0.007, N=4) than the Hausdorff distance for the acute NPV, ranging from 6.09 to 12.14 mm (8.34 $\pm$ 2.28 mm).

\section{Discussion}

The rigorous spatial evaluation of the acute and post-NPV MR biomarkers against the gold-standard histology label is only possible because of the presented MR to histology registration method.
This evaluation revealed that NPV biomarker acquired 3-5 days after treatment was a more accurate predictor of the MRgFUS induced tissue necrosis than the acute NPV.
Although this result is consistent with prior studies that have used qualitative or indirect metrics to compare NPV and histology \cite{payne2013vivo, futureneedsHectors2016}, the presented results quantitatively demonstrate the significant difference between the two metrics.
The acute NPV over predicted the histology label for every subject, as can be seen in \figureautorefname~\ref{fig:volume_comp}, reinforced by the low precision score, high recall score, and larger volume. 
Looking at the contour comparisons in \figureautorefname~\ref{fig:volume_comp}, the acute NPVs for subjects 2, 3, and 4 have distinct features that are not present in the histology label. 
These large, mislabeled regions are reflected in the large Hausdorff distances between the acute NPV and histology, with the most obvious mislabeled structure in subject 4 with a Hausdorff distance of 12.14 mm.  
These regions and general overestimation of the acute NPV could be a result of transient effects previously described in the literature, such as vascular occlusion \cite{futureneedsHectors2016}.
These results demonstrate the acute NPV biomarker measured on CE-T1w MR imaging is not an accurate predictor of histological tissue viability following MRgFUS ablation for the presented model. 

The post-NPV MR biomarker significantly outperformed the acute NPV in every spatial metric except for recall.
This result is expected because the overestimation yields a high recall score, which explains the lack of a significant difference between the acute NPV and the post-NPV recall. 
However, the mean precision of the post-NPV biomarker was significantly higher, indicating that the post-NPV biomarker had fewer false positives compared to the histology label than the acute NPV. 
The qualitative alignment of the post-NPV to histology volume in \figureautorefname~\ref{fig:volume_comp} is quantitatively reflected in the significantly lower and more consistent reported Hausdorff distances relative to the acute NPV. 
It is important to note that these spatial metric results cannot be used to evaluate the accuracy of the registration because the post-NPV MR biomarker may not perfectly match the histology volume even with perfect registration. 
These correlations between \textit{in vivo} MR and histology are what we aim to understand and therefore cannot be used to drive or evaluate the registration without biasing the final result. 

The average TRE of 1.00 $\pm$ 0.13 mm resulting from the presented registration workflow is lower than the TRE reported by state-of-the-art MR to histology registration methods used in similar applications \cite{rusu2020registration}.
The TRE of the presented registration workflow is on the order or the resolution of several clinical MR biomarkers, making the workflow suitable for evaluating these biomarkers \cite{futureneedsHectors2016, schmitz2010precise}. 
Additionally, the reported registration accuracy is achieved without relying on features that correlate directly between \textit{in vivo} MR images and histology that are independent of the diagnostic or treatment features being evaluated.
R3 registration between \textit{in vivo} MR and \textit{ex vivo} MR in the presented method may include treatment features. 
Correlating features between \textit{in vivo} and \textit{ex vivo} MR is possible because the image contrast is consistent, and the deformation is minimal between the two stages. 
However, the direct relationship between \textit{in vivo} and the H\&E stained tissue sections is unknown, so features between \textit{in vivo} MR and histology may or may not directly correlate.
Therefore, selecting features directly between \textit{in vivo} MR and histology for registration would bias the evaluation of MR biomarkers against the histological response.
Additionally, evaluating the registration from landmarks directly between \textit{in vivo} MR and histology might not accurately evaluate the accuracy of registration because the selected features may not correlate. 
To address this problem, we use different landmarks for each stage for registration evaluation as the correlation between two stages is better defined. 
Although our method may include treatment features, we minimize the dependence to one stage and only between images of the same MR modality. 
Using intermediate imaging makes the presented registration workflow applicable for not only improving registration in prior applications, but also validating of a wider range of applications, including MRgFUS ablations in the liver, breast, or kidneys.

A direct comparison between registration methods is not possible due to large differences in the overall workflows and differences in the included anatomies. 
Prior registration methods often rely on a 2D correlation between \textit{in vivo} MR and histology  \cite{li2017co,pichat2018survey,losnegaard2018intensity,iglesias2018joint,orczyk2013preliminary,dickinson2013image,goubran2015registration, rusu2020registration}. 
However, this study emulated the 2D slice correspondence assumption in our \textit{ex vivo} registration step and compared the error of the presented method against this assumption to demonstrate the advantage of the 3D block reconstruction method. 
Prior studies have shown that this 2D slice correlation assumption limits the accuracy of the registration to a minimum of $\sim$ 1mm \cite{gibson20133d}.
We find similar limitations to this assumption in our results. 
For our pipeline, the 2D slice assumption introduces significant amounts of error into the registration pipeline, as shown in \figureautorefname~\ref{fig:stage_tre}. 
Although the  X and Y dimension error could be improved, the amount of unrecoverable error introduced in the Z dimension using the 2D assumption is still significantly (p = 0.015) more than the total error using our novel block reconstruction method. 
The high registration accuracy, novel block reconstruction method, and independence of correlating MR and histology features clearly show the advantages of the presented multi-stage registration workflow.  

The presented registration workflow and metric evaluation do not come without limitations. 
The time required for tissue processing and data reconstruction is not trivial. 
During microtome sectioning (D3), sequentially sectioning through each tissue block while acquiring digital blockface images every 50 $\mu m$ is currently time consuming. 
In this work, each tissue block must be fully sectioned in order to obtain a 3D model for the \textit{ex vivo} registration step (R2). 
However, there are some methods to generate 3D models of the tissue blocks that are more automated and less time intensive, such as micro-CT. 
Micro-CT can be used to extract the 3D surface of each tissue block without sectioning, leaving the amount of microtome sectioning and blockface imaging up to the discretion of the user. 
The postprocessing relies on semi-automatic segmentation of blockface imaging and histology sections, which is time consuming at the high resolutions needed to generate high-fidelity models. 
However, the neural networks used to provide initial segmentations in this study will become more accurate with additional samples, leading to more automatic processing. 
Although the presented registration workflow requires more time compared to other workflows, the improved registration accuracy results in a reduction of the necessary number of samples to effectively evaluate MR biomarkers \cite{gibson20133d, rusu2020registration, pichat2018survey}.
Although the small sample size limits the statistical significance of the NPV evaluation, this work demonstrates how the presented registration workflow can be used to rigorously evaluate any kind of MR biomarkers. 
Future studies utilizing the presented registration method can be added to these results to improve the statistical power of the biomarker evaluation. 
Additionally, as more data becomes available, current and novel biomarkers can be prospectively studied to fully understand their accuracy for predicting histology necrosis. 

Despite these limitations, the presented registration workflow facilitated extensive comparisons between the commonly utilized MRgFUS NPV biomarker and gold-standard histology for viability.
Contextualize the meaning of the precision and recall scores for MRgFUS ablation treatments of oncological targets is important.
If an MR biomarker has low precision with high recall, such as the acute NPV biomarker, a large number target pixels may be labeled as treated, although they are still viable. 
Relying solely on these MR biomarkers may result in untreated cancer remaining after MRgFUS ablation.
Although the acute NPV biomarker is not optimal for immediate MRgFUS treatment assessment, several promising acute MR biomarkers will need to be evaluated against histology, including T2w imaging, MR temperature imaging, and diffusion imaging \cite{hectors2014multiparametric,plata2015feasibility,mannelli2009assessment,haider2008dynamic,wu2006registration}. 
The presented registration methods can facilitate evaluating and validating these proposed biomarkers in future studies. 

We presented a novel, multi-step MR to histology registration workflow that corrects all deformations without assuming any feature correlation between \textit{in vivo} MR and histology, but rather between intermediate imaging steps.
The accuracy of the presented registration workflow has a lower error than previously reported state-of-the-art \textit{in vivo} MR to histology registration methods.
The novel contribution of our presented workflow is the block reconstruction step depicted in \figureautorefname~\ref{fig:block_recon}.
The presented registration workflow facilitated novel spatially accurate evaluation of MRgFUS NPV biomarkers against the gold-standard label of tissue viability and can be used to evaluate and develop novel, acute \textit{in vivo} MR metrics that can more accurately predict the tissue viability measured with histology. 
Validating \textit{in vivo} MR biomarkers will increase their clinical viability for MRgFUS treatment assessment and facilitate clinical translation of minimally invasive treatments for oncological applications.

\section{Methods}

The presented MR to histopathology registration pipeline was evaluated in a large animal tumor model.
Supplemental Video 1 provides an animation of the entire tissue processing and reconstruction workflow.  
To evaluate both the acute NPV and the post-NPV as MR imaging biomarkers, the acute NPV was registered to the post-NPV using methods described in Zimmerman \textit{et al.} \cite{zimmerman2020learning}
The tissue from the subject was excised immediately after follow-up imaging according to the procedures outline in \sectionautorefname~\ref{sec:extraction}-\ref{sec:microtome}.
The histology was reconstructed and registered to the follow-up \textit{in vivo} MR images according to the procedures outlined in \sectionautorefname~\ref{sec:blockfacereg}-\ref{sec:invivoreg}. 

\subsection{Subject Model}
\label{sec:model}
A VX2 cell suspension (1\texttimes10\textsuperscript{6} cells in 50\% media/Matrigel) was bilaterally injected intra-muscularly into the quadriceps of four New Zealand white rabbits and grown for 1-2 weeks. 
Anesthesia was induced with a ketamine/xylazine injection (IM, 25/5 mg/kg), and the animal was then intubated, allowing anesthesia to be maintained with inhaled isoflurane (0.5-4.0\%) for the duration of the MRgFUS treatment. 
Animal vitals, including temperature and respiration, were monitored throughout treatment. 
Hair on the treated quadriceps was removed via clippers and a depilatory cream (Nair\textsuperscript{\texttrademark}) to enable acoustic coupling. 
Using a pre-clinical MRgFUS system (Image Guided Therapy, Inc.), ablation was performed on one tumor and the surrounding quadriceps muscle tissue with a 256-element phased-array transducer (Imasonic, Voray-sur-l'Ognon, France; 10-cm focal length, 14.4 x 9.8 cm aperture, f=940 kHz) inside a 3T MRI scanner (PrismaFIT Siemens, Erlangen, Germany). 
Ablation details for the four subjects are outlined in \tableautorefname~\ref{tab:acoustics}.
A single loop MR receiver coil was incorporated into the MRgFUS system table to improve the image signal to noise ratio (SNR) around the targeted quadriceps.  
The ablation procedure was monitored in real time with MR thermometry imaging (MRTI) using 3D-segmented echo planar imaging sequences.
Following ablation therapy, the animal was recovered and monitored for 3 days. 
After 3 days, the animal was re-anesthetized and follow-up (post) imaging was performed.
For more details on the \textit{in vivo} MR imaging, we refer the reader to Zimmerman \textit{et al.}\cite{zimmerman2020learning}
The animal was euthanized immediately following imaging. 
The University of Utah Institutional Animal Care and Use Committee (IACUC) approved all procedures (\#17-08012, September 7, 2017).
All methods and procedures were performed in accordance with the IACUC guidelines and regulations.

\begin{table}[ht]
    \ra{1.1}
  \centering
  \caption{Ablation details for each subject including time between treatment MR (where the acute NPV is measured) and follow-up imaging (where post-NPV is measured), ablation points, acoustic power output, and total energy achieved for each subject. }
  \small
  \begin{tabular}{ccccc}
    \toprule
    & \multicolumn{1}{c}{Follow-up} & \multicolumn{1}{c}{Number of} & \multicolumn{1}{c}{Acoustic Power} &  \multicolumn{1}{c}{Total}\\
  \#. & {Duration} & {Sonications} & {Mean $\pm$ 1 Std. (W)} & {Energy (kJ)} \\
    \midrule
    1& \cellcolor{gray!40}5 days &11 & \cellcolor{gray!40}57 $\pm$ 17 & 23.14 \\
    2& \cellcolor{gray!40}3 days&12 & \cellcolor{gray!40}69 $\pm$ 18 & 26.00 \\
    3&\cellcolor{gray!40} 5 days&14& \cellcolor{gray!40}44 $\pm$ 9  & 18.59 \\
    4& \cellcolor{gray!40}5 days&10 & \cellcolor{gray!40}56 $\pm$ 9  & 18.55 \\
    \bottomrule
  \end{tabular}\vspace{-4pt}
  \label{tab:acoustics}
\end{table}\vspace{-5mm}

\subsection{D1: Tissue Excision}
\label{sec:extraction}
The treated quadriceps was surgically excised immediately following euthanization.
Pathology inks were applied to the tissue during excision to maintain the MR orientation, and the excised tissue was submerged in 10\% formalin solution for 14 days. 
The fixed \textit{ex vivo} tissue was mounted in a custom tissue-processing box and encapsulated in 3.5\% agar solution for \textit{ex vivo} imaging and subsequent gross slicing. 
The custom box and pathology inks facilitated orienting the tissue as close to the \textit{in vivo} MR orientation as possible.
The tissue-processing box incorporated a single-loop MR coil around the box to improve \textit{ex vivo} MR image SNR.
T1w and T2w MR images of the agar-embedded tissue were acquired using a 3T MRI scanner (PrismaFIT Siemens, Erlangen, Germany).
The field of view and voxel size for \textit{in vivo} and \textit{ex vivo} T1w MR images used in the registration pipeline was 256 x 56 x 192 mm with 0.5 x 1.0 x 0.5 mm spacing, and for T2w MR the field of view and spacing was 256 x 52 x 192 with $1.0$ mm isotropic spacing.

\subsection{D2: Gross Slicing}
\label{sec:slicing}
Following \textit{ex vivo} imaging, the agar embedded tissue block was grossly sliced along the head-foot axis of the \textit{ex vivo} MR imaging in $\sim$3 mm increments using a deli slicer (Backyard Pro SL110E). 
The surrounding agar was removed from each sliced tissue block without disturbing the tissue blocks.
The exposed tissue faces from gross slicing were re-inked with pathology inks to indicate the head and foot surfaces and maintain the orientation of each block relative to the original \textit{ex vivo} sample. 
Each tissue block was placed into an individual whole-mount tissue cassette for further formalin fixation (5 days). 
After the additional fixation, each block was embedded in paraffin wax with consistent orientation.

\subsection{D3: Microtome Sectioning}
\label{sec:microtome}

Each paraffin wax tissue block was sectioned at 10 $\mu$m increments with a microtome (Leica RM2255, Leica Microsystems, Wetzlar, Germany).
Digital images of the blockface were acquired every 50 $\mu$m starting from the very beginning of the block using a digital single-lens reflex (DSLR) camera (Nikon D7100; Macro 1:1 105mm Lens; 2.0x teleconverter).
The image size and approximate resolution for a blockface image are $6000 x 4000$ with $\sim 0.018$ mm isotropic spacing.
Paraffin wax is slightly transparent, so tissue from behind the exposed face would show through on the images. 
To address this, two images were acquired at every 50 $\mu$m with two different lighting conditions. 
The first image had the light approximately aligned with the camera whereas the second image had the light perpendicular to the camera. 
Taking the difference between two lighting conditions shows only the tissue that is exposed on the surface.
These blockface images were acquired automatically with an Arduino board controlled with a Python GUI. 
The Python GUI automatically triggered the camera and lights, transferred files to the computer, recorded image information (name, section depth, camera settings, etc), and backed up each image to cloud storage to ensure data retention. 
Three 5 $\mu$m thick tissue sections were retained on glass slides for future histopathology analysis every 250 $\mu$m.
Sequential sampling was repeated until there was no tissue remaining in the paraffin wax. 
After sectioning, the first section of each group of three was stained with hematoxylin and eosin (H\&E).
The remaining sections were reserved for additional future stains. 
All H\&E stained sections were imaged with a brightfield microscope (Axio Scan, Zeiss, Oberkochen, Germany) at 2.5 magnification. 
The native resolution of the microscopic images is $\sim 0.0076 mm$, but the images were down-sampled to the resolution of the blockface images for use in registration.  
Labels of necrotic tissue on the H\&E stained sections were semi-automatically generated using Gaussian mixture modeling and expert manual segmentation.

\subsection{R1: Histology to Block Registration}
\label{sec:blockfacereg}

Each histopathology section was registered to the nearest incremental 50 $\mu$m blockface image.
The digital histology images were down-sampled to the resolution of the blockface image ($\sim$ 0.02$\times$0.02$\times$0.05 mm).
An affine transform between the histology and blockface was solved via automatic intensity-based affine registration of the blockface segmentation and the histology segmentation.
Following affine registration, a multi-scale, intensity-based registration was used to deformably register the histopathology segmentation images to the corresponding blockface image segmentations. 
Each image was registered by minimizing the following energy $E$: \vspace{-1pt}
\begin{equation}
    E = \int_{\Omega} \norm{I_1(\varphi^{-1}(\vec x, t)) - I_0(\vec x)}^2,
    \label{eqn:image_energy}
\end{equation}\vspace{-1pt}
where $I_1$ is the histopathology segmentation,  $I_0$ is the blockface image segmentation, $\varphi^{-1}(\cdot)$ is a diffeomorphism, and $\Omega$ is the image domain.
\equationautorefname\:\ref{eqn:image_energy} was optimized using a gradient flow algorithm with a Cauchy-Navier operator \cite{christensen1996deformable}.
This registration step provided a diffeomorphism between a histology section and its corresponding blockface image that corrected deformations introduced during microtome sectioning.

\subsection{R2: Tissue Blocks to \textit{Ex Vivo} Registration}
\label{sec:blockrecon}

Surface registration techniques were used to drive the restoration of the blocks to their original morphology and account for deformations from gross slicing. 
A 3D surface, represented by a triangular mesh object, was constructed for each tissue block and the \textit{ex vivo} tissue from their segmentations using a marching cubes algorithm. 
The exterior tissue surface was segmented from \textit{ex vivo} MR to generate a 3D surface of the \textit{ex vivo} tissue that was used as the target for block reconstruction. 
For each tissue block, the 2D blockface images were stacked to create a 3D blockface volume. 
Automatic intensity-based 2D affine registration was used to register sequential blockface images to account for small shifts in camera position during imaging.
Finally, blockface volumes were semi-automatically segmented using a custom-trained 3D V-Net neural network and manual segmentation to generate the 3D surfaces of each tissue block \cite{milletari2016v}.

Each tissue block was represented by a surface, $S_b$ where $b$ indicates the block number ranging from one to the number of total blocks from the gross slicing, $nb$.
For a single tissue block, the mesh was semi-automatically separated using surface normals and manual segmentation into a head surface $S_b^h$, foot surface $S_b^f$, and exterior surface $S_b^e$, such that $S_b = S_b^h\cup S_b^f\cup S_b^e$.
Examples of the corresponding faces used for registration are shown in \figureautorefname~\ref{fig:block_recon} (b).  
The corresponding surfaces were registered together using surface-based registration, which is outlined in \sectionautorefname\:\ref{sec:surfaceReg}.
The algorithm in \figureautorefname\:\ref{alg:reconblocks} provides an overview of the block reconstruction to \textit{ex vivo} process. 
$d(\cdot, \cdot)$ refers to the surface distance metric defined in \sectionautorefname\:\ref{sec:surfaceReg}.
In general, a center block was chosen as the starting point for the reconstruction, and deformations were solved for block by block, propagating outwards from the center block first in the head direction, then in the foot direction.
Once the deformation was determined for one block, the deformed block then became the target of registration for the next sequential block.
Each block had an associated affine transformation and diffeomorphism, $A_b$ and $\varphi^{-1}_b$, respectively. 
After all blocks were registered and stacked back together, the deformed exterior surfaces from each block were joined together and registered with the \textit{ex vivo} tissue surface.  

\begin{figure}[ht]
\centering
\footnotesize \small
\begin{minipage}{0.47\textwidth}
    \begin{spacing}{1.05}
	\centering
	\begin{algorithmic}[1]
		\Procedure{Rebuild Blocks}{$\{S_1 \dots S_{nb}\}$}
		\State Center Block for Target $S_t$where $t=nb \; // \; 2$
		\For{\texttt{i = 1; i $\leq$ nb // 2; i++}}
		\State b $=$ (nb // 2) - i
        \State $\hat{S}_b^f = \min_{A_b} d(\:S_t^h, \: A_b \circ S_b^f \:)$
        \State $\bar{S}_b^f = \min_{\varphi^{-1}_b} d(\:S_b^h, \: \varphi^{-1}_b \circ \hat{S}_b^f \:)$
        \State $S_t^h = \varphi^{-1}_b \circ A_b \circ S_b^h$
        \EndFor
        \State Reset $S_t$ to center block $t=nb \; // \; 2$
		\For{\texttt{i = 1; i $\leq$ nb // 2; i++}}
		\State b $=$ (nb // 2) + i
        \State $\hat{S}_b^h = \min_{A_b} d(\:S_t^f, \: A_b \circ S_b^h \:)$
        \State $\bar{S}_b^h = \min_{\varphi^{-1}_b} d(\:S_b^f, \: \varphi^{-1}_b \circ \bar{S}_b^h \:)$
        \State $S_t^f = \varphi^{-1}_b \circ A_b \circ S_b^f$
        \EndFor
        \State Union deformed exterior surfaces $S^e = \bigcup \{\bar{S}_b^e; b \in np \}$
        \State $\hat{S}^e = \min_{A} d(\:S_{\textit{ex vivo}}, \: A \circ S^e \:)$
        \State $\bar{S}^e = \min_{\varphi^{-1}} d(\:S_{\textit{ex vivo}}, \: \varphi^{-1} \circ \hat{S}^e \:)$
		\EndProcedure
	\end{algorithmic}
	\end{spacing}
    \end{minipage}
    \vspace{-20pt}
	\caption{Algorithm overview for reconstruction of the tissue blocks to the \textit{ex vivo} tissue surface. \vspace{-7mm}}
	\label{alg:reconblocks}
\end{figure}

\subsection{R3: \textit{Ex Vivo} to \textit{In Vivo} Registration}
\label{sec:invivoreg}

Expert segmentations of corresponding features, such as blood vessels, tumor, or treatment features, identified from the T2w images from the \textit{ex vivo} and \textit{in vivo} MR data were used to generate 3D surfaces in both spaces. 
An affine transform and diffeomorphism between \textit{ex vivo} and \textit{in vivo} MR was solved using 3D surface-based registration defined in \sectionautorefname\:\ref{sec:surfaceReg}.
The full transformation from the \textit{in vivo} to \textit{ex vivo} environment was comprised of an affine matrix and a diffeomorphism.
All deformations from the three restoring registration pipeline steps were composed to yield a final diffeomorphism that was a spatial mapping between \textit{in vivo} MR and any histopathology section. 

\subsection{Surface-Based Registration}
\label{sec:surfaceReg}

Surface-based registration was used throughout the reconstruction pipeline to solve for affine transforms and diffeomorphisms between two 3D surfaces. 
The surfaces did not have corresponding landmarks that could be identified from the images or surfaces. 
We implemented a previously developed unlabeled point-set matching algorithm to solve for a diffeomorphism between the two surfaces \cite{glaunes2004diffeomorphic}.
All surface objects were defined as triangular mesh objects. 
For a single triangular mesh object $S$, the kernel norm (K-norm) is defined as \vspace{-1pt}
\begin{equation}
    \norm{\sum_{p=1}^{n \in S} \eta(p) \delta_{c(p)}}_K^2 = \sum_{p=1}^{n} \sum_{p'=1}^{n} \; \langle \eta(p), \eta(p') \rangle \; K(c(p),c(p'))
\end{equation}\vspace{-1pt}
where $p$ is a triangular element of $S$, $n \in S$ is the number of faces in surface $S$, $\eta(p)$ the normal vector to $p$, $c(p)$ is the center point of triangle $p$, $K(\cdot, \cdot) = k(\cdot, \cdot)I$ where $I$ is a $3\times3$ identity matrix and $k(\cdot, \cdot)$ is a scalar valued Cauchy kernel. 
The dissimilarity between two surfaces $S_1$, $S_2$ is driven by the difference between the sum of the vector valued Dirac masses centered at the triangle center with \vspace{-1pt}
\begin{equation}
\begin{split}
    \norm{\sum_{p=1}^{n \in S_1} \eta(p) \delta_{c(p)} - \sum_{q=1}^{m \in S_2} \eta(q) \delta_{c(q)}}_K^2 =  &\sum_{p=1}^{n} \sum_{p'=1}^{n} \; \langle \eta(p), \eta(p') \rangle \; K(c(p),c(p')) - 
    2  &\sum_{p=1}^{n} \sum_{q=1}^{m} \; \langle \eta(p), \eta(q) \rangle \; K(c(p),c(q))  \\  + 
     &\sum_{q=1}^{m} \sum_{q'=1}^{m} \; \langle \eta(q), \eta(q') \rangle \; K(c(q),c(q')).
\end{split}
\end{equation}\vspace{-1pt}
For a more detailed explanation on how affine transformations and diffeomorphisms act on these surfaces objects, we refer the reader to Glaunes \textit{et al} \cite{glaunes2004diffeomorphic}.

\subsection{NPV Biomarker Analysis}
After performing registration, the acute NPV biomarker, post-NPV biomarker, and the histology necrosis label were co-registered in the follow-up MR imaging space (where post-NPV was measured).
The spatial accuracy of the acute NPV and the post-NPV biomarkers were evaluated against the histology necrosis label using precision, recall, DICE coefficient, and Hausdorff distance.  
The acute NPV vs histology and post-NPV vs histology spatial metrics were compared using two-sample-independent t-tests. 

\subsection{Registration Accuracy}
Landmarks were used to validate the accuracy of the registration via the target registration error (TRE).
Selecting landmarks directly between MR and histology is difficult for two reasons: first, selecting landmarks between 2D and 3D spaces is challenging, and the relationship between MR treatment features and histology treatment features is being evaluated and cannot be used to evaluate the registration. 
Consequently, we selected three sets of N=20 landmarks (five per subject) for each of the four subjects to evaluate each step of the registration pipeline individually.
The TRE was calculated via the Euclidean distance between the deformed landmarks and the target landmarks for each stage of registration. 
The upper bound of the total registration pipeline was estimated by summing the mean error from each individual stage. 

The novel contribution of our registration pipeline is the 3D block reconstruction to insert each histology slice into the 3D \textit{ex vivo} MR space. 
Prior registration methods assumed a 2D relationship between 2D histology and a 2D MR slice.
We emulate this assumption during our 3D block reconstruction by restricting the Z dimension (perpendicular to the histology sections) block motion to only translation, effectively assuming that the 2D histology slices align with a 2D \textit{ex vivo} MR slice. 
Using the landmarks, we calculate the TRE under this 2D projection assumption to demonstrate the improved accuracy of the presented 3D block reconstruction. 
The TREs of the two different methods were compared using two-sample-related t-tests on the Euclidean error between the deformed landmarks and target landmarks.

\bibliography{ms}

\end{document}